\documentclass[preprintnumbers,superscriptaddress,preprint]{revtex4-1}
\usepackage{textcomp}
\usepackage{amsmath}
\usepackage{amssymb}
\usepackage{graphicx}
\usepackage{color}
\usepackage{soul}
\newcommand{\ntu}
{\affiliation{School of Science and Technology, Nottingham Trent University, Clifton Lane, Nottingham NG11 8NS, UK}}

\newcommand{\mpi}
{\affiliation{Max Planck Institute for Dynamics and Self-Organization, Am Fassberg 17, 37077 G\"{o}ttingen, Germany}}

\begin{document}

\title{Mapping heterogeneities through avalanche statistics}

 \author{Soumyajyoti Biswas}
 \email{soumyajyoti.biswas@ds.mpg.de}
 \mpi
 
 \author{Lucas Goehring}
 \email{lucas.goehring@ntu.ac.uk}
 \ntu
 \date{\today}

 \begin{abstract}
 Avalanche statistics of various threshold activated dynamical systems are known to depend on the magnitude of the drive, or stress, on the system. Such dependencies exist for earthquake size distributions, in sheared granular avalanches, laboratory scale fracture and also in the outage statistics of power grids. In this work we model threshold-activated avalanche dynamics and investigate the time required to detect local variations in the ability of model elements to bear stress.  We show that the detection time follows a scaling law where the scaling exponents depend on whether the feature that is sought is either weaker, or stronger, than its surroundings.  We then look at earthquake data from Sumatra and California, demonstrate the trade-off between the spatial resolution of a map of earthquake exponents (\textit{i.e.} the $b$-values of the Gutenberg-Richter law) and the accuracy of those exponents, and suggest a means to maximise both. 
\end{abstract}
%\pacs{64.60.F-,75.10.Nr,64.70.Tg,75.50.Lk}
\maketitle

\section{Introduction}
Catastrophic breakdowns can occur as events in widely disparate situations, from laboratory scale fracture, to the collapse of buildings or bridges, outages in power grids or high magnitude earthquakes. However, there is a common theme binding together these phenomena, which occur at various length and energy scales: they all represent threshold-activated intermittent dynamics, powered by a relatively slow external drive and having some form of internal dissipation \cite{REVIEW}.  The past decades of research in these fields have revealed the connection between the statistics of their intermittent responses and the physics of self-organised criticality (SOC) \cite{ref1}. In a self-organised critical system, external drive, for example slow ($\sim$mm/year) tectonic movements for earthquakes, and dissipation can conspire to generate an attractive fixed point at the critical state of the system. Consequently, large-scale correlations develop, which are used to explain the universal nature of these phenomena of vastly different origins. 

An unwelcome consequence of the critical nature of these problems is that their system-wide catastrophic failure is much more likely than if they were far from criticality. Particularly, for SOC the size distributions of avalanche events are known to scale as $P(S)\sim S^{-B}$, where $S$ is the size of any rapid avalanche of activity and $P(S)$ is the cumulative size distribution of those avalanches. For earthquakes, this is known as the Gutenberg-Richter law \cite{Gutenberg, explanation}, and similar laws are also valid for laboratory scale fracture \cite{ref2}, outages in power grids \cite{ref3} \textit{etc.}   Much of the damage caused by intermittent failure events is naturally associated with the extreme tail of their size distributions \textit{i.e.} the large-scale events such as devastating earthquakes or nation-wide blackouts. Therefore, considerable research effort has been spent on finding ways to forecast such major events, from the time series of any system activity that precedes them.  
 
One such series of efforts has led to the awareness that the exact value of the exponent, $B$, depends on the magnitude of the forcing on the system \cite{ref4}. For example, in the case of earthquakes the distribution of seismic activity in regions of higher tectonic stress shows a lower $B$ value than occurs on average \cite{ref5,scholz2015,gori,singh,tormann}.  Since a lower $B$ value translates into a higher risk of larger events, \textit{i.e.} a higher risk of large-magnitude earthquakes, several authors have compiled maps of $B$ values for their use as statistical hazard assessments (\textit{e.g.} see References \cite{ref9,gori,tormann,singh}).  Similar stress dependencies are also observed for sheared granular media \cite{ref6}, the collapse of cliffs \cite{amitrano} and in power grid outages \cite{ref7}.  

A common difficulty in this forecasting process, however, lies in the resolution and extent of the data.  Resolution limits can be due to the accuracy of the instruments used to locate an avalanche, or they may be due to choices made in how the data was stored.  For example, corrections are required when delving deep into the historical earthquake record, as detection equipment has evolved over time \cite{handbook}.  Alternatively, for power outages, public data reports are often limited to the resolution of the nearest city \cite{USdata}.   This scale may be small compared to the total spread of the event across the power grid, but remains large compared to the individual locations of failed lines. Another limitation is in the number of observations that can be collected, as often large numbers of events need to be analysed before any significant statement about the size distribution exponent, $B$, can be made. The spatial and temporal limitations in making a risk-map are also intimately connected. While a larger number of observations can more accurately determine the $B$ value relevant to any particular location, for any finite data set this will create the need to draw these data from a larger sampling region, thus resulting in a loss of spatial resolution.

In this work we investigate how much data is required to detect a local variation in the stress profile of a disordered system. Within a simple discrete model, we initially study the number of observations required to uniquely distinguish one part, or region, that has a different relative load from the rest of the system. We arrive at the counterintuitive result that the scaling of the number of events required for such distinction is different depending on whether the part considered is stronger or weaker than its surrounding, even by the same magnitude.  We then look at how the measured exponent, $B$, varies with distance away from an embedded inhomogeneity.  Finally, using earthquake data for the Sumatra region, we demonstrate how the uncertainty of a $B$-value measurement depends only weakly on the number of events used to determine it.  We use this result to explore the interplay between the spatial resolution of a $B$-value map, and its accuracy.  

\section{Model}

\begin{figure}
\centering\includegraphics[width=4.5in]{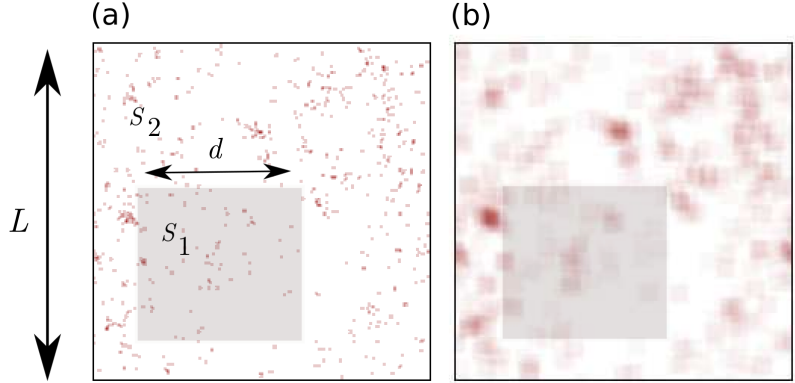}
\caption{Model schematic, showing the distribution of failure locations in a single example avalanche event.  (a) A system of size $L$ is randomly loaded.  Most elements have an average excess capacity to bear load of $s = s_2$, while elements on a sub-region of size $d$ are assigned a different average excess capacity $s = s_1$.  This embedded region can thus be stronger ($s_1 > s_2$) or weaker ($s_1 < s_2$) than its surroundings.  The scatter of red points shows the locations of failed elements, for an avalanche triggered by the failure of a single randomly chosen site.  (b) We convolve the resulting failure activity with a window of size $\ell$, to reflect any observational uncertainties.  This blurs the avalanche pattern, as in the example here.  The results are then split into grid cells, also of size $d$, and the avalanche size distribution in each grid region is measured separately.}
\label{fig_config}
\end{figure}

The model used here is similar to that in Ref.\cite{ref7}, which we use to model the dynamics of power outages, and which is, in turn, adapted from a class of similar models \cite{ref8,yagan1,yagan2}. Specifically, we consider a set of $L \times L$ elements arranged in a square lattice. Every element is assigned a load, $\sigma_l$, and a failure threshold, or a maximum capacity to bear load, $\sigma_{th}$.  The simulation starts with a stable configuration given by 
\begin{equation}
\sigma_{th}^i=\sigma_l^i+\epsilon^i s,
\label{stable}
\end{equation}
where $s$ represents the average excess capacity of the system to bear load, $\sigma_l^i$ is the load on the $i$th element and $\epsilon^i$ is a random variable. The individual values of $\sigma_l$ and $\epsilon$ are both taken from a uniform distribution in the range $[0,1]$. To start the dynamics, a single randomly chosen element is broken, \textit{i.e.} its threshold is set to zero, and its entire load is redistributed to the rest of the system, such that an element at a Euclidean distance $r$ (assuming periodic boundary conditions) receives a fraction proportional to $1/r^2$.  This long-range redistribution is used to reflect the long-range nature of the interactions present in many systems, including in earthquakes \cite{eq_long}, fracture \cite{REVIEW} and power grids  \cite{power_long}. Specifically, the inverse square law reflects the stress field near an Inglis crack within an elastic solid \cite{sadd}, or the electric field near a defect (or crack) embedded in a conducting material \cite{alava}.  The redistribution can raise other elements above their thresholds, such that they also fail: if so, their loads are then redistributed to their own surroundings, in turn.   This process continues until the avalanche of failures ends by either reaching a stable configuration, or by breaking all the elements on the lattice.  The long-range nature of the load redistribution means that the individual failures in an avalanche do not need to be spatially connected.  Indeed, this is known to be the case in the remote triggering of both earthquakes \cite{remote} and power outages  \cite{96outage}. The number of elements breaking, before the system stabilises, determines of the size of one avalanche.   Every element in the lattice is then refreshed and reset with new randomly chosen variables, satisfying Eq. \ref{stable}, ready to simulate another avalanche.

In the present version we make two modifications to this model. First, in order to simulate places of relative weakness or strength, we allow the value of $s$ to depend on location. Specifically, we set $s = s_2$ everywhere, except for a particular region, of size $d\times d$, where $s = s_1$.  A lower value of $s$ means that the system is closer to its threshold, so a weak or damaged area can be simulated by setting $s_1<s_2$.  The opposite condition, $s_1>s_2$, would represent a particularly strong section of the system.  This is a way of formulating stress heterogeneity in the system. We note that in some other models, including the Burridge-Knopoff and Olami-Feder-Christensen models, nucleation centres are dynamically generated (see \textit{e.g.} \cite{kawamura1,kawamura2}). In the present context, however, we introduce an asperity like behaviour in the construction of the model itself.   

The second modification involves how the avalanche locations are recorded. Here, we blur the location of each failed element by convolving the final distribution of failure events with a square mask of size $\ell\times\ell$. This process is illustrated in Fig. \ref{fig_config}, and is intended to reflect any observational uncertainties of the activity in a system.  Therefore, with this model we can investigate how long it takes to detect any spatial heterogeneities in a near-critical set of elements, given an imprecise knowledge of the avalanche locations. 

 The data are typically analysed by rank plots, where events $S$ are arranged in descending order of their size, so that $ S_1 \ge S_2 \ge S_3 \dots \ge S_n$.  The $k$th ranked element has size $S_k$.  For events with a probability distribution $p(S)$, the number of events having size greater than or equal to $S_k$ is now
\begin{equation}
  \int_{S_k}^{\infty} p(S)dS=k.
\end{equation}
However, the left hand side of this equation is a cumulative distribution that is assumed to follow a power-law, $P(S)\sim S^{-B}$ (dropping the index $k$). Therefore, if the events follow a self-similar distribution, then $ k\sim  S_k^{-B}$, and this relationship will be used to determine the exponent $B$~{\cite{newmann05}.

\section{Results}

For earthquakes there is a negative correlation between the scaling exponent of the rank-plot of earthquake magnitudes, and the stress in the region under study \cite{ref5,scholz2015,explanation}.   We can capture this motivating result in our model by setting $s_1 = s_2$, so that nothing distinguishes different regions of the system.  The avalanche-size distribution of this homogeneous system now depends only on its relative load and capacity.   Figure~\ref{newfig}(a) shows such distributions for four such cases, in a system of size $L = 200$.  Here, a higher relative stress is equivalent to a smaller excess capacity, or smaller value of $s$.  We find that the rank-plots of the system-wide activity show a similar dependence to those of earthquakes, and decrease in slope with decreasing $s$, representing increasing stress.   As shown in Fig.~\ref{newfig}(b), the value of $B$ varies continuously with $s$, until the critical point $s = 0.5$ is reached. This critical point was empirically determined as the average capacity below which a systemwide failure becomes the favoured result following a single broken element, as also occurs in \textit{e.g.} Ref. \cite{pradhan}. 

\begin{figure}
\centering\includegraphics[width=130mm]{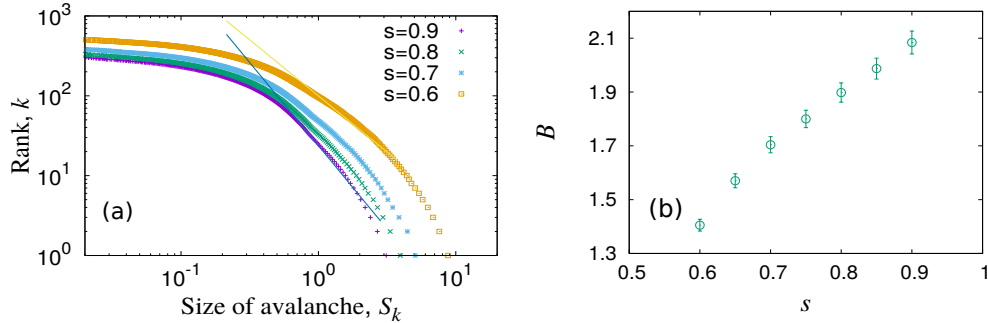}
\caption{Avalanche statistics for uniform systems, where $s_1 = s_2 = s$, show scale-free behaviour.  (a) When arranged in rank by size, the distribution of avalanches shows a power-law tail.  For small events, the blurring of the event location leads to a systematic flattening of the distribution.  (b) By fitting a power-law to the tail of these distributions, we extract the exponent $B$, which decreases in magnitude as the critical point of $s = 0.5$ is approached.  A similar, roughly linear, dependence of $B$ on local levels of stress is known to occur for earthquakes \cite{ref5,scholz2015}.}
\label{newfig}
\end{figure}

\subsection{Temporal resolution}

\begin{figure}
\centering\includegraphics[width=135mm]{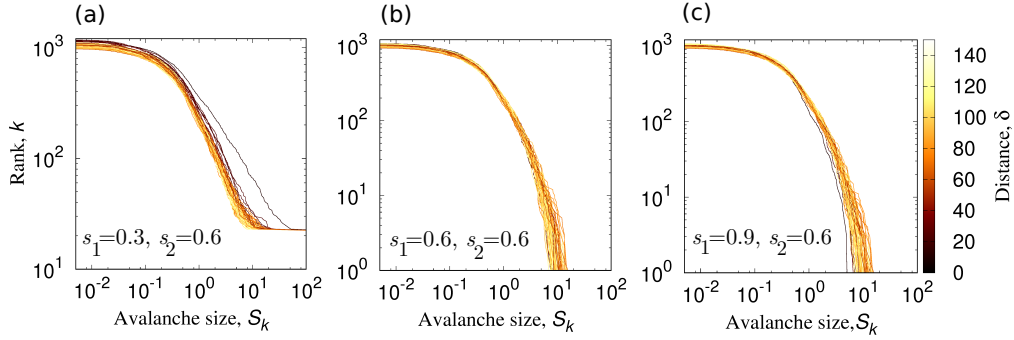}
\caption{Avalanche size distributions for all the gridded regions are shown, after 10 000 events. The colours indicate the average distance of each grid cell to the centre of the embedded region. (a) For a weaker or damaged inclusion, the avalanche size distribution there is clearly shallower than elsewhere.  However, nearby regions, indicated by darker lines, also show lower exponent values than the average.  (b) If $s_1 = s_2$ there is no distinction between regions, and a random spread in distributions is seen, due only to random sampling errors.  (c) For a stronger inclusion, the scaling exponent on that grid cell is noticeably higher than elsewhere. The neighbouring regions are also affected by the inclusion.}
\label{all_expo_space}
\end{figure}

The first question that we address is how long it will take to unambiguously identify a patch of weaker or stronger material, embedded in a larger system, by analysing the statistics of avalanche events.  We consider a lattice of $200 \times 200$ elements, containing a block of size  $d = 20$  on which $s = s_1$, and where $s = s_2$ elsewhere.  The strength of the local inhomogeneity is thus $\Delta s = s_2 - s_1$, such that $\Delta s<0$ represents a strong patch, for example.  For simplicity, and to avoid introducing multiple independent length scales into the analysis, we also mask the resulting activity with a window of size  $\ell = 20$.  The whole system is then divided up into 100 different grid cells containing $20 \times 20$ elements, and the scaling coefficient $B$ is calculated for the activity on each different grid cell, by a least-squares fit to the rank-size distribution.  Similar results are also found if the maximum likelihood estimator (MLE) method \cite{MLEClauset} is used.  

The limitation in the resolution (both spatial and temporal) comes from the facts that (i) the locations of individual avalanches are imprecisely known, as the the exact locations are convolved with a mask of size $\ell$, and (ii) the system is divided into several (in this case 100) observation boxes. It is important to note that the observation of a ``sub-sample" within the whole sample is not the same as a finite size scaling analysis \cite{viola}. Particularly in cases such as the present one, where the coupling (load redistribution range) is long-range, a part of the whole system will be strongly affected by the regions surrounding it.

In Fig. \ref{all_expo_space} we show three sets of results, where $s_2 = 0.6$, and $s_1 =0.3, 0.6,$ and $0.9$.   In other words, situations where the embedded region is respectively weaker than, identical to and stronger than the rest of the system.  Note that the blurring and subdivision of the activity data means that the minimum size of an event that can be detected in a region is only $1/d^2 = 0.0025$.  In each case 10 000 avalanche events are monitored.  The figure shows the natural spread of avalanche size distributions, due to the finite sample size.   However, the lines there are coloured to represent the relative positions of the different observation windows.  The darkest line represents the distinct patch, on which $s = s_1$, and the colours get lighter as the grid cells get further away from this region.  It can be seen that, due to the long-range nature of the stress redistribution rule, a large area surrounding the embedded region gets affected by its different relative strength.  Specifically, when $s_1\neq s_2$, a gradient in the colour of the lines can be seen to be associated with the variation in the slopes of those lines.   This effect will be explored further in Section 3(b).

\begin{figure}
\centering\includegraphics[width=5.5in]{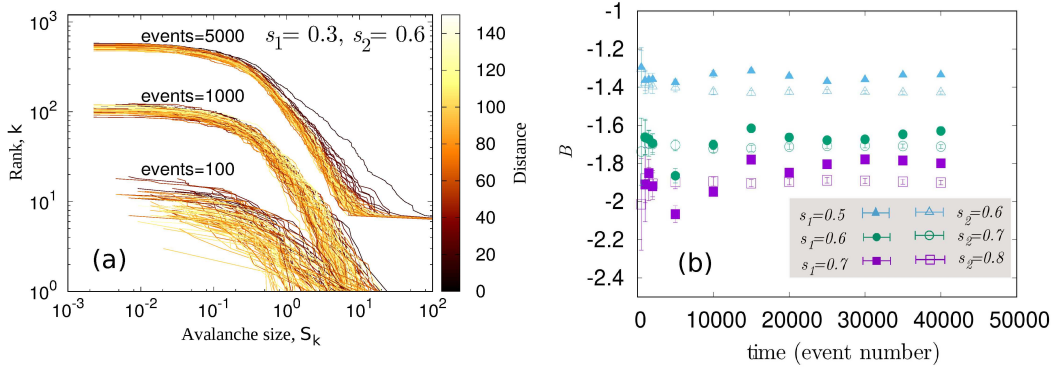}
\caption{The temporal evolution of the avalanche size distributions is shown. (a) Although the relative loads on the defect and elsewhere are very different, for small numbers of avalanches (\textit{e.g.} 100 or even 1000) such distinctions are not reflected in the avalanche size distributions. Only after 5000 steps can the exponent at the defect location be clearly discerned from its neighbours. (b) The measured exponent values at the defect and the rest of the system are shown with time, for three different situations.   While initially it is not possible to distinguish the defect region, at later times the distinction becomes clear, supporting the graphical measure shown in (a).}
\label{all_expo_time}
\end{figure}

Thus, for a long enough observation period and a large enough difference, $\Delta s$, the exponent values of different regions are clearly distinguishable.  Such distinctions are harder to observe when the number of observed avalanches, or events, is few.   This is demonstrated in Fig. \ref{all_expo_time}(a), where the same distributions are shown after 100, 1000, and 5000 avalanches have been recorded.   In spite of a large difference between the relative loads, \textit{i.e.} $\Delta s = 0.3$, little clear difference can be seen in the regional avalanche size distributions, up to observations of about 1000 events.  This is relevant for analysing real data sets, as we will demonstrate in Section 4 with the Sumatra earthquake catalogue. In other words, it can take a large number of observations before even a relatively prominent defect cannot be reliably located, as the noise of near-critical fluctuations can introduce false signals.

To study the question of detection time more clearly, we varied the value of $s_1$ for a given background $s_2$ and measured the time required for unique detection. This is taken to be the earliest time, \textit{i.e.} number of avalanches, beyond which the average avalanche size in the modified grid cell is consistently the largest or smallest in the entire system, depending on whether that region is weaker or stronger than its surroundings.   In other words, it represents the number of observations required before the activity in the embedded patch will be able to stand out from its surroundings.  This definition of detection time will depend weakly on system size, and will diverge when the difference between the average strength of the defect and the rest of the system vanishes.   However, it is also consistent with other related metrics, such as the number of events required before the measured $B$ value of the embedded region is measurably distinct from its surroundings (see Fig.~\ref{all_expo_time}(b)).

\begin{figure}
\centering\includegraphics[width=4.0in]{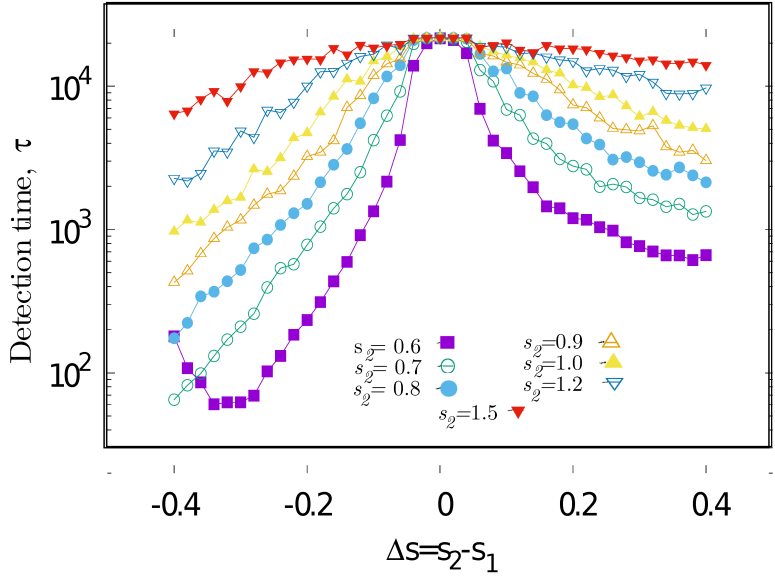}
\caption{The time required to uniquely identify the location of a heterogeneity is plotted as a function of the relative variation of its sites, $\Delta s$, and the excess capacity of its surroundings, $s_2$.  Generally, the closer the system is to criticality, the more activity there is, and the easier detection becomes.  This is captured by the different scaling behaviours for the detection time, which is not symmetric around $\Delta s = 0$.  While for an infinite system the detection time should diverge as $\Delta s$ approaches zero, for a finite system and observation time this simply converges to a large value.}
\label{detection_time}
\end{figure}

Figure \ref{detection_time} shows how the detection time of our system depends on the difference between the average strengths of the embedded region, and the rest of the system, \textit{i.e.} $\Delta s$.   To further analyse the behaviour of the detection time, we note that the response on either side of $\Delta s=0$ shows two different exponents, and is not symmetric.  Specifically, a scaling is seen of the form
\begin{equation}
\tau \sim \mathcal{F}\left(\frac{\Delta s}{s_2^{\alpha}} \right),
\label{collapse}
\end{equation} 
where
\begin{equation}
 \mathcal{F}(x)\sim x^{-\theta}.
\end{equation}
For small $|\Delta s|$ the detection time saturates due to finite system size.  However, for a large enough system, and sampling time, it should naturally diverge as $\Delta s$ approaches zero.  Interestingly, we can now see that the exponents $\theta$ and $\alpha$ depend on the sign of $\Delta s$, \textit{i.e.} whether the detection time is for a weaker patch on a stronger background or a stronger patch on a weaker background. This is demonstrated in Fig. \ref{expo_time_collapse}, which shows the collapse of the two branches of detection time results, for different exponent choices.

\begin{figure}
\centering 
\includegraphics[height=5cm]{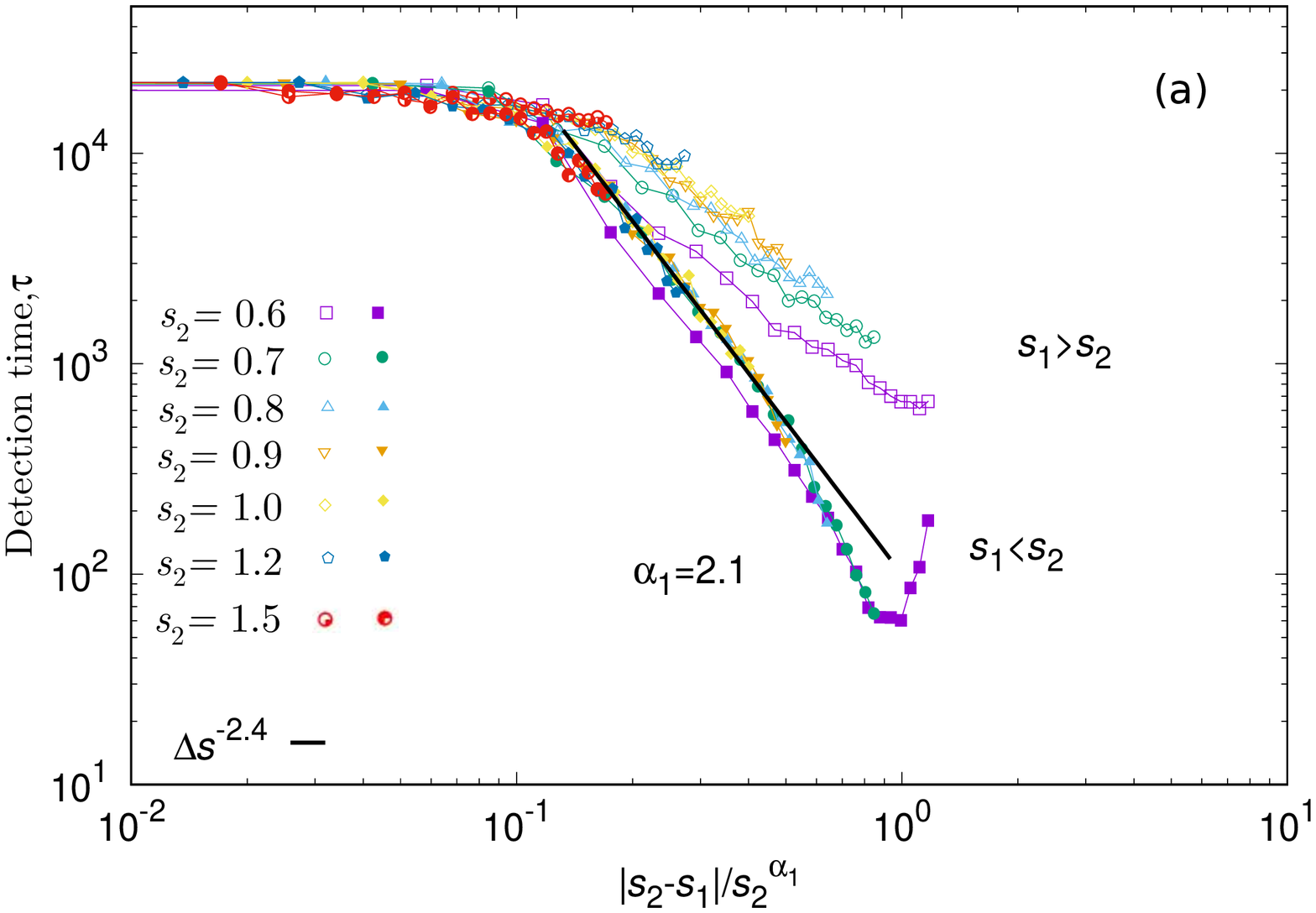}
\includegraphics[height=5cm]{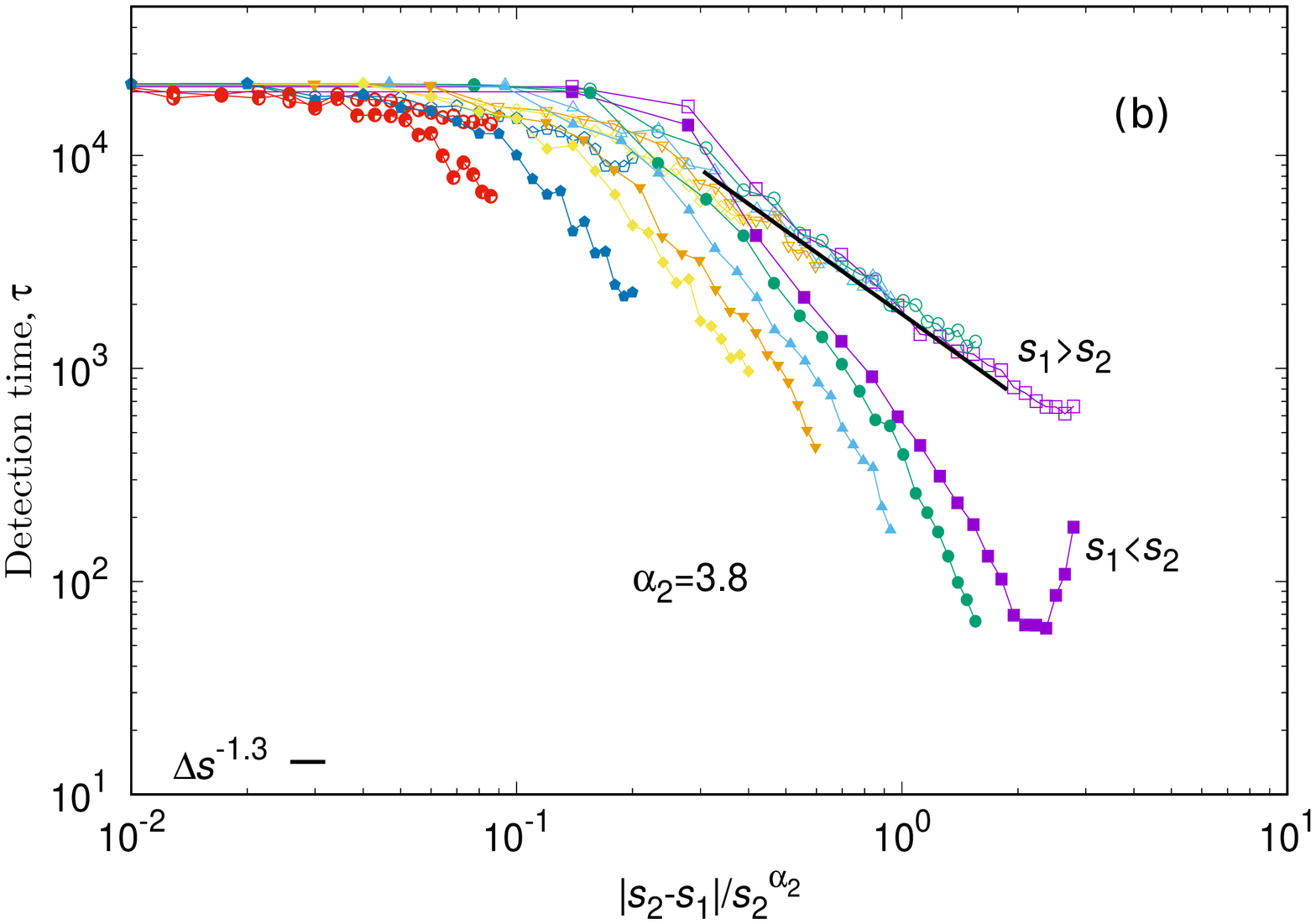}
   \caption{The detection time collapses onto two master curves, depending on whether the embedded region is (a) weaker, or (b) stronger than its surroundings.  The data collapse is done in both cases following the form of Eq. \ref{collapse}. The values for the set of the exponents $\alpha, \theta$ are different for the two cases $s_1<s_2$ and $s_1>s_2$ and the two branches are related \textit{via} the relationship given in Eq. \ref{scaling} to within the measured numerical accuracy.}
\label{expo_time_collapse}
\end{figure}

While an analytical estimate for $\theta$ and $\alpha$ appears challenging, there is a scaling relation between the sets of exponents on either side of $\Delta s$. The detection times on either side of $\Delta s=0$ should converge to the same value for small $\Delta s$. This is because, when $s_1$ and $s_2$ are very close, the probabilities that the average avalanche size at the defect is lower or higher than average are equal (the whole system is homogeneous). By extension, therefore, the probability and the time-scale for which the defect is detected as the weakest or the strongest are also equal. This implies that for a small enough magnitude of $x$,
\begin{equation}
\tau(x)=\tau(-x),
\end{equation}
giving
\begin{equation}
\frac{\Delta {s_0}^{-\theta_1}}{s_2^{-\alpha_1\theta_1}}=\frac{\Delta {s_0}^{-\theta_2}}{s_2^{-\alpha_2\theta_2}},
\end{equation} 
where $\Delta s_0$ is a small constant.  This limit must remain valid for all choices of $s_2$, which can only happen when 
\begin{equation}
\alpha_1\theta_1=\alpha_2\theta_2.
\label{scaling}
\end{equation}
From the fits to the data in Fig. \ref{expo_time_collapse}, we estimate $\alpha_1=2.1\pm0.05$, $\theta_1=2.4\pm0.05$, $\alpha_2=3.8\pm0.05$, and $\theta_2=1.3\pm0.05$, giving $\alpha_1\theta_1=5.04\pm0.23$ and $\alpha_2\theta_2=4.94\pm0.26$. Given the quality of the data, the scaling relation is valid within the numerical accuracy. 

\subsection{Spatial resolution}
Given sufficient time, the avalanche activity in inherently stronger or weaker region will allow it to be detected.  Due to the correlated, and long-range nature of avalanches, this time can be surprisingly long, requiring analysis of several thousands of events to come to an unambiguous conclusion, which is often not possible with data from various real situations, for example, power outages.  Other than insufficient observation time, the detection of a defect location  is 
dependent on the spatial extent and relative strength of the defect with respect to that of the observation grid. It is also dependent on the 
location of the observation window with respect to the actual defect. 
 Here, in preparation for looking at a real earthquake record, we will briefly consider  the spatial distribution of events around the embedded region.

%In Fig. \ref{expo_ratio_size_time} these two 
%effects are studied. Particularly,
%the ratio $B_1/\langle B_2\rangle$, where $B_1$ is the exponent on the grid containing the defect and $\langle B_2\rangle$
%is the average value of the exponents on all other grids, decreases as the defect size is increased. This is primarily due to the
%decrease in the exponent value at the defect location.  On the other hand, for a given defect size, the exponent ratio varies almost linearly with
%the ratio of the relative loads on the defect and the rest of the system. This is due to the linear variation of the
%exponent with the relative load.

\begin{figure}
\centering
\includegraphics[width=3.0in]{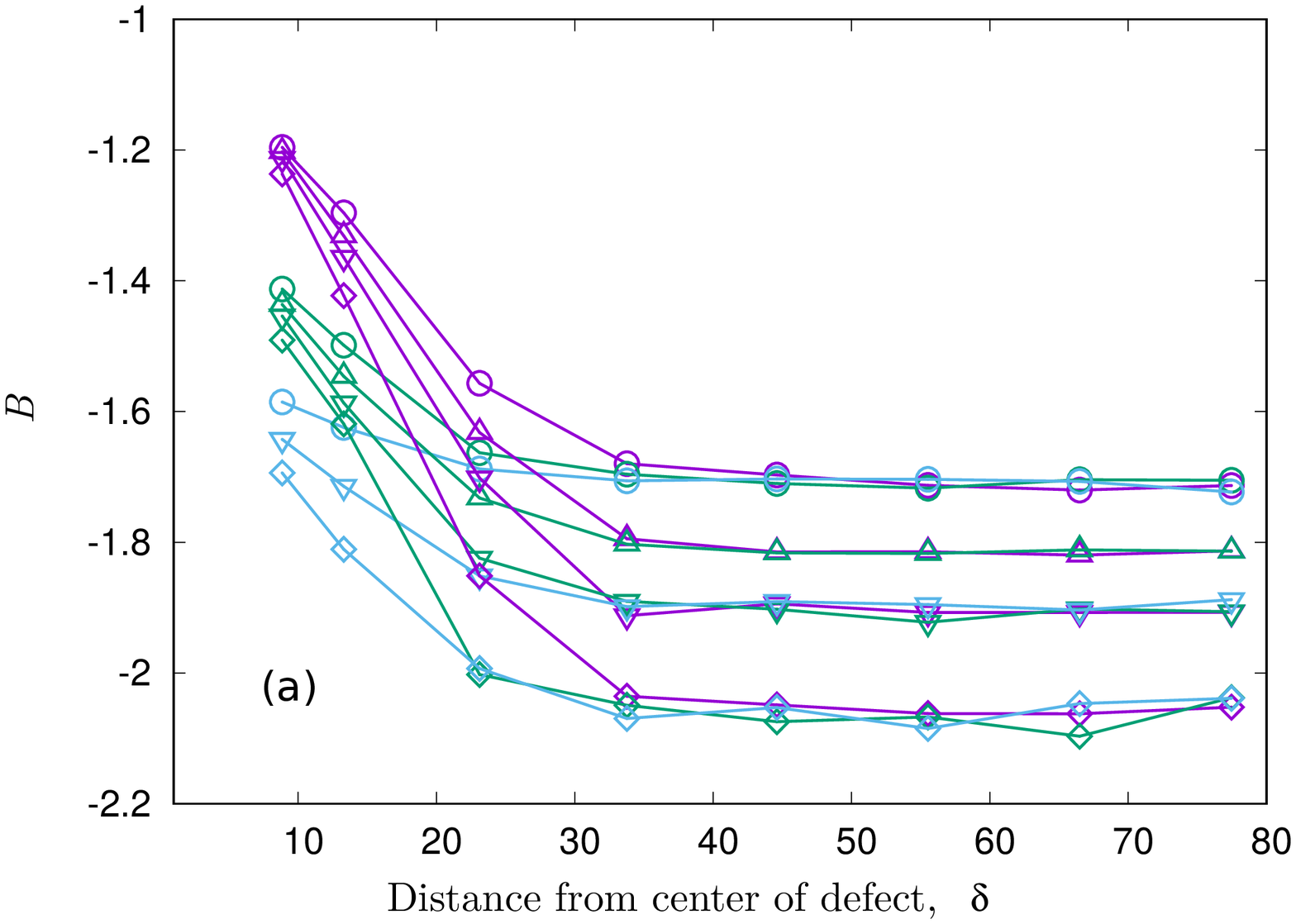}
\includegraphics[width=3.0in]{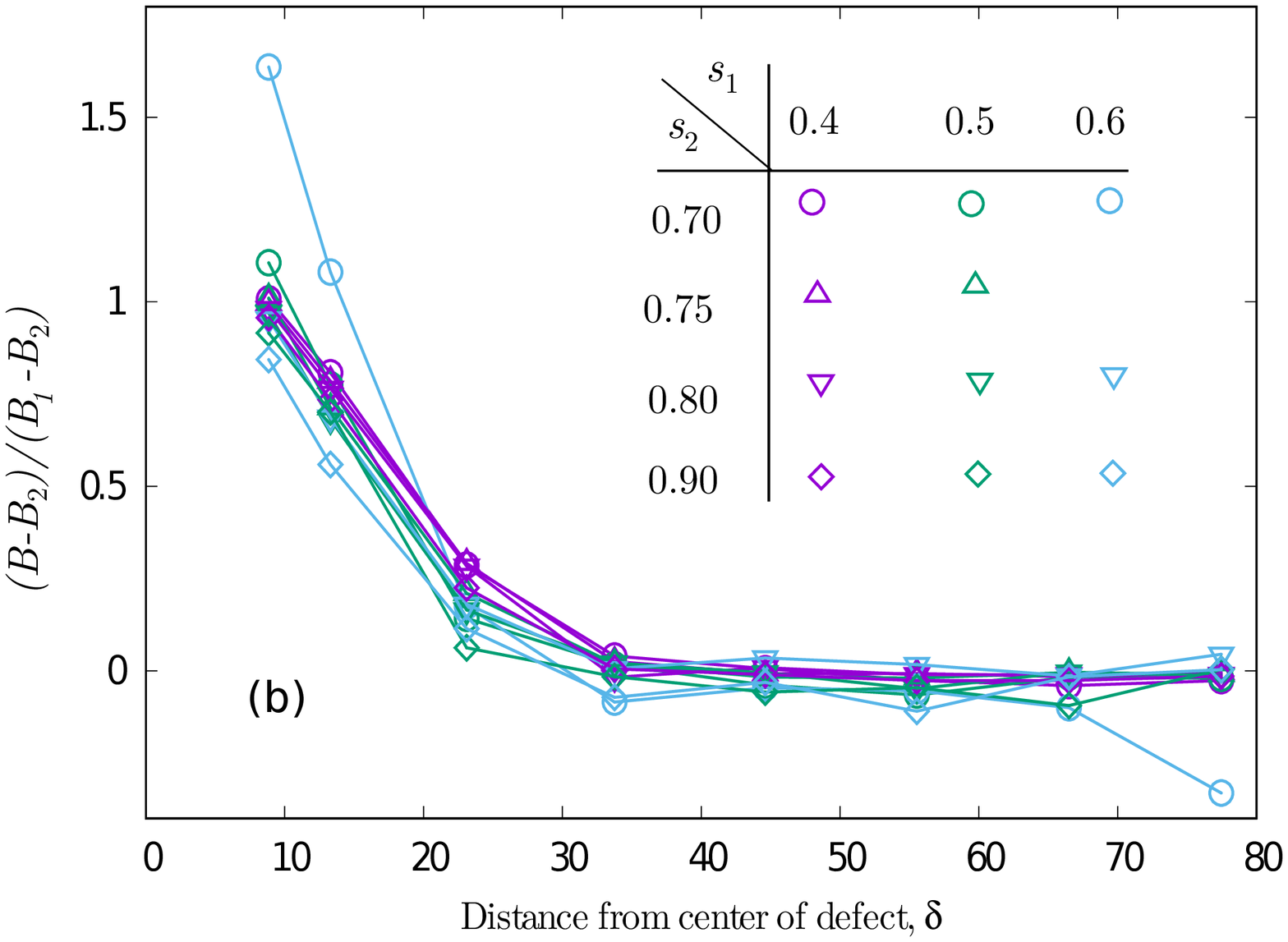}
%%% where xxxxxx name represents "figurename.eps"
\caption{The spatial variations in avalanche size exponents are shown.  Panel (a) shows how the exponent of the avalanche size distribution is enhanced in a wide region around an inclusion with a different relative strength.  (b) The curves for different combinations of $s_1$ and $s_2$ values collapse onto one master curve when their variations are scaled by the difference of the exponents in the limiting cases (see Eq. \ref{fd}).  In other words, long-range correlations in avalanche behaviour means that the effects of embedding a region of distinct strength can be felt relatively far away from that region.}
\label{expo_distance_new1}
\end{figure}

%Also, as we have noted before, the exponent values in different grid points are affected by their proximity to the defect site
%as well. 
In Fig. \ref{expo_distance_new1} we plot the exponent value noted in one grid point as it is shifted away from the location 
of the defect. The figure at the bottom shows a function of the form

\begin{equation}
F(\delta)=\frac{B-B_2}{B_1-B_2},
\label{fd}
\end{equation}
where $B_1$ and $B_2$ are the exponent value when the grid coincides with the defect and the grid is far away from the 
defect, respectively. A collapse of the curves show that the variation of the exponent from a lower to a higher magnitude
depends only on $\Delta s$, when we have kept $s_1<s_2$. As mentioned in Section 3(a), due to the long range load redistribution
rule of the model, the exponent values in the region surrounding the defect, where the relative strength is higher, are also
affected by the presence of the defect. The decay of the function $F(\delta)$ sets the detection scale of a defect embedded in 
a long range interacting system.   

\subsection{Earthquake data}
Finally, we look at how temporal and spatial resolution affects the mapping of variations in stress levels in real earthquake data. For this we choose (i) the 11628 events in the USGS earthquake catalogue  from the Sumatra region, between 10$^\circ$S to 15$^\circ$N latitude and 90$^\circ$E to 100$^\circ$E longitude, in the period from 2000 through 2010, inclusive,  and (ii) the 9987 events in the USGS catalogue from the California region (32-42$^\circ$N and 117-125$^\circ$W) for the same period of time.   The first case gives a comparable data set to that studied in Ref. \cite{ref9}, which demonstrated regional variations in the scaling exponent, $b$, of the Gutenburg-Richter (GR) law, or magnitude-frequency distribution, of earthquakes there. It is worth mentioning that this data also contains the magnitude 9.1 event on December 26th, 2004. In fact, the fifteen largest earthquakes occurring in the data set all show a lowering of the $b$ value, as reported in \cite{ref9}.  As earthquake magnitude is reported in terms of energy, we note that $b = 2B/3$ \cite{explanation}.  Similar studies have been performed on different scales and across the globe, including around the L'Aquila fault in Italy \cite{gori}, in California \cite{tormann} and in the NW Himalayas \cite{singh}.  

Our measurement procedure for the $b$ value is similar to that used in Ref. \cite{ref9}. For each case, we impose a grid of size $0.25^\circ\times0.25^\circ$ on the whole region. Centering on each of the grid points, a growing circular region was considered until a given number of events, $n$, was captured within. For large $n$ this circle may be considerably larger than the grid resolution, and its size will also vary with event frequency.  The exponent value at each grid point was then found by fitting the magnitude-frequency distribution of those $n$ closest events, using a least-squares fitting method (\textit{n.b.} as in section 3(a), similar results were also obtained for MLE fits).   The error on the fit is also recorded as $\varepsilon$.  Evidently, if a larger $n$ is chosen, then there are more data points to fit to the GR law and hence a lower error estimate for $b$.  On the other hand, to get a higher $n$ the radius of the observation window has to be larger.  This then limits the spatial resolution of any possible $b$ value map of the region.  Finally, as earthquake statistics are generally correlated in both space and time, different events cannot truly be considered independent.  This will affect the rate at which uncertainties in $b$ can be expected to decrease with the sample size.  

In Fig. \ref{expo_distance_product}(a) we show how the mean fitting errors, $\bar\varepsilon$, averaged over all the grid points along the Sumatra subduction zone as well as the California region, change with the number of events used to fit each exponent.  It shows a power-law dependence on $n$, with an exponent value of $-0.27\pm0.01$ for the Sumatra region and $-0.25\pm0.01$ for the California region.  This is considerably shallower than the relationship expected for independent random variables, namely $\varepsilon \sim n^{-0.5}$.  However, this relationship need not hold for observations having correlation, and some results to this effect are known from the analysis of GPS data \cite{mao,behr}.  Obviously, for larger $n$ the average radius $\bar d$ of the observation window also increases.  To maximise the information content of a spatial risk map, or heat map, of seismic activity, one would aim to minimise the product $\bar\varepsilon\bar d$.  We therefore plot this metric in Fig. \ref{expo_distance_product}(b), as a function of $n$.   It remains almost constant up to $n\approx 100$, with a shallow dip near $n=50$ and a significant increase above about $n=100$ for the Sumatra region, and the for the California region, the constant part is somewhat extended.   In Fig. \ref{expo_map} we demonstrate this trade-off between the accuracy of the $b$-values and their spatial resolution, by showing maps calculated with different values of $n$ for the Sumatra subduction zone.
 
\begin{figure}
\centering
\includegraphics[width=135mm]{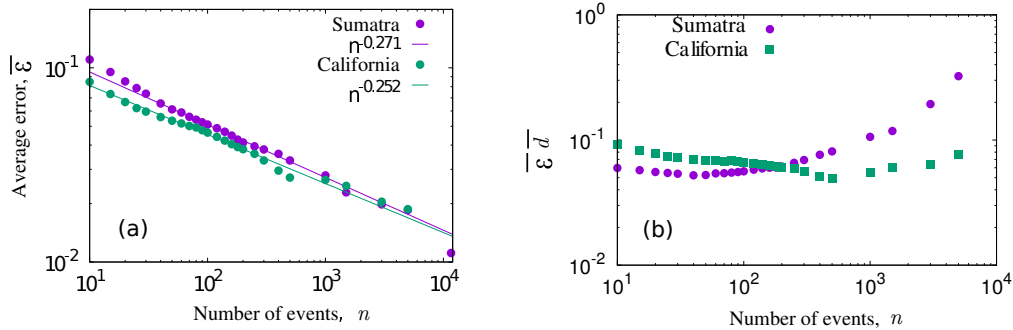}
\caption{Optimising uncertainty in $b$-value determination.  (a) The average error in fitting the $b$-values of the exponent of the Gutenberg-Richter law is plotted against the number of events considered for that estimate, $n$, for the earthquake data covering the Sumatra subduction zone and California region. For several decades the error decreases as a power law with an exponent of $-0.27\pm0.01$ for Sumatra and with $-0.25\pm0.01$ in California. This is considerably shallower than would be the case for random, uncorrelated events.  The average error, in the limit $n\to 11628$ tends to the value of $\pm0.01$ noted in Ref. \cite{ref9} for the entire Sumatra data set.  (b) The product of the average error, $\bar\varepsilon(n)$ and the average distance from the centre each grid point required to find $n$ events, $\bar d(n)$, is plotted as a function of $n$, and represents the total uncertainty in each measurement. The variation for the Sumatra region is more or less constant up to $n=100$, and beyond that it increases rapidly, but has a weak minimum around $n=50$. For the California region, the variation is qualitatively similar, while the constant part of the curve is somewhat more extended.} 
\label{expo_distance_product}
\end{figure}

\begin{figure}
\centering
\includegraphics[width=135mm]{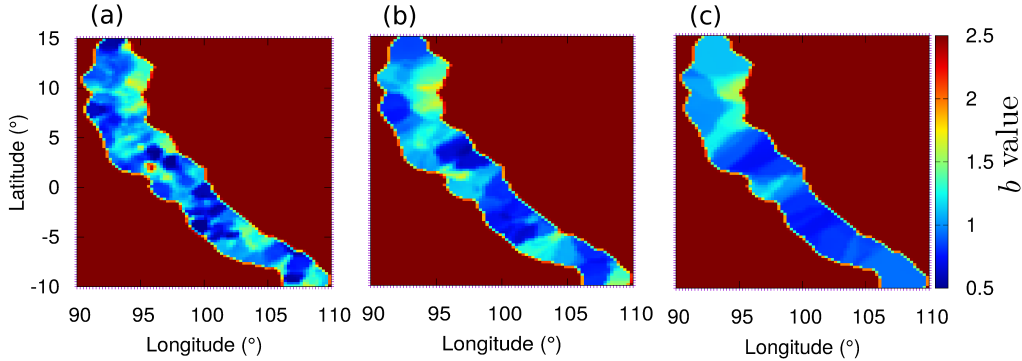}
\caption{Maps of the fitted $b$-values in the region of the Sumatra subduction zone are shown for different choices of $n$, the number of events used to measure the $b$ value at a given grid point. Shown are the results for (a) $n = 50$, (b) $200$ and (c) $800$ events.  As $n$ increases, the effective spatial resolution in the maps drops, although the observed features are more reliably reported.  The map with $n=50$ optimises the information displayed in the figure, by minimising the uncertainty in the product $\bar\varepsilon\bar d$.}
\label{expo_map}
\end{figure}

\section{Conclusion}
Variations in the exponent values that characterise the relative sizes of events in a driven disordered system provide a useful method to forecast vulnerable regions of that system. This method is relevant to evaluating risk in diverse situations, including earthquakes \cite{ref5,scholz2015}, power outages \cite{ref7} and cliff failure \cite{amitrano}.  Some challenges in applying it, however, are the limited size of some of the most important data sets, such as earthquake or power outage records, and any inherent observational resolution of that data.  These limitations must be taken into account in the construction of any possible risk-maps, as they affect the ability to forecast imminent large events. 

In this work we have shown, using a discrete and minimal model for failure dynamics, that the spatial and temporal resolutions of the rank-plot exponents are interrelated.  By embedding a region which differed from its surroundings in its capacity to bear load, we investigated how many events needed to be recorded before we could unambiguously find that distinct patch of the system.  We found that the timescale required for detection followed power-laws that depended on whether the embedded region was stronger or weaker than its surroundings.  However, we also described how the scaling exponents of the two cases are related, and demonstrated this \textit{via} data collapse.  We then showed
that spatial variations in the measured exponents across the system can extend well beyond the borders of any distinct area, or inhomogeneity in the system's excess capacity to carry stress.  Again, we demonstrated this \textit{via} data collapse, in this case by interpolating between the exponent values expected in the embedded region, and in areas far from it.

Finally, we have shown the implications of this coupling of spatial and temporal resolutions on the measurement of
the $b$-values of the Gutenberg-Richter law.  For this, we chose to map the earthquake data of the Sumatra subduction zone, using different numbers of events to estimate the scaling exponent of each map point.   The accuracy of the $b$ values was found to depend only weakly on the number of events used in each fit, presumably due to correlations between the events.  As a result, we found that the product of the spatial resolution of such a fit, and the error in determining its $b$ values, can be used to produce a risk-map that attempts to maximise the accuracy of the information presented.
%\vskip6pt

%\enlargethispage{20pt}

\section*{Data Access:}
The data for the earthquake events in the Sumatra region are taken from the USGS 
website: https://earthquake.usgs.gov/earthquakes/search/.
\section*{Author contributions:}
Both authors were involved in all aspects of the paper.
\section*{Acknowledgements}
S.B. acknowledges the Alexander von Humboldt Foundation for support.
%\competing{The author(s) declare that they have no competing interests.}

%\acknowledgement{}

%\ack{Insert acknowledgment text here.}


\begin{thebibliography}{100}

\bibitem{REVIEW}
H. J. Herrmann, S. Roux (eds.), {\it Statistical models for the
fracture of disordered media}, Elsevier (North-Holland), 1990.

\bibitem{ref1} P. Bak, C. Tang, K. Wiesenfeld, \textit{Self-organized criticality: An explanation of the 1/f noise}, Phys. Rev. Lett. {\bf 59}, 381 (1987)

\bibitem{Gutenberg}
G. Gutenberg, C. F. Richter, {\it Earthquake magnitude, intensity, energy and acceleration},
Bull. Seismol. Soc. Am. {\bf 46}, 105 (1956).

\bibitem{explanation}
Note that the $B$ values for earthquake statistics when measured in terms of the exponent of the size distributions of the energy emitted in an event is related to
the Gutenberg-Ricter exponent $b$, where the magnitudes of the events are measured, by $b=2B/3$, see I. G. Main, L. Li, J. McCloskey, M. Naylor, Nat. Geosci. {\bf 1}, 142 (2008).  

\bibitem{ref2} S. Biswas, P. Ray, B. K. Chakrabarti, (2015)  \textit{Statistical physics of fracture, breakdown and earthquakes}.
 Wiley-VCH, Berlin. 

\bibitem{ref3} B. A. Carreras, D. E. Newman, I. Dobson, \textit{Evidence for self-organized criticality in electric power system blackouts}, IEEE Trans. Circuits Syst. {\bf 51}, 1733 (2004).

\bibitem{ref4} C. H. Scholz, \textit{The frequency-magnitude relation of microfracturing in rock and its relation to earthquakes}, Bull. Seism. Soc. Am. {\bf 58}, 399 (1968).

\bibitem{ref5} D. Schorlemmer, S. Wiemer, M. Wyss, \textit{Variations in earthquake-size disribution across different stress regimes}, Nature {\bf 437}, 539 (2005).

\bibitem{scholz2015}
C. H. Scholz, {\it On the stress dependence of the earthquake {\it b} value},
Geophys. Res. Lett. {\bf 42}, 1399 (2015).

\bibitem{gori}
P. de Gori, F. Lucente, A. Lombardi, C. Chiarabba, C. Montouri,
{\it Heterogeneities along the 2009 L'Aquila normal fault inferred by the 
b-value distribution}, Geophys. Res. Lett. {\bf 39}, L15304 (2012).

\bibitem{singh}
C. Singh, {\it Spatial variation of seismic {\it b}-value across the
NW Himalaya}, Geomatics, Natural Hazards and Risk {\bf 7}, 522 (2014).

\bibitem{tormann}
T. Tormann, S. Wiemer, A. Mignan, {\it Systematic survey of high-resolution 
{\it b} value imaging along Californian faults: Interface
on asperities}, J. Geophs. Res. Solid Earth {\bf 119},
2029 (2014).

\bibitem{ref9} P. Nuannin, O. Kulhanek, L. Persson, \textit{Variations of b-value preceding large earthquakes in the Andaman-Sumatra subduction zone}, J. Asian Earth Sci. {\bf 61}, 237 (2012).

\bibitem{ref6} T. Hatano, C. Narteau, P. Shebalin, \textit{Common dependence on stress for the statistics of granular avalanches and earthquakes}, Sci. Rep. {\bf 5}, 12280 (2015).

\bibitem{amitrano}
D. Amitrano, J. R. Grasso, G. Senfaute, {\it Seismic precursory patterns before a cliff collapse and critical point phenomena}, Geophys. Res. Lett. {\bf 32}, L08314 (2005).

\bibitem{ref7} S. Biswas, L. Goehring, \textit{Load dependence of power outage statistics}, arxiv:1805.07792v1 (2018).

\bibitem{handbook}
W. Lee, P. Jennings, C. Kisslinger, H. Kanamori (eds.), {\it 
International handbook of earthquake \& engineering seismology (part A)},
Academic press, 2002.

\bibitem{USdata}
see, for example, outage data for the U.S. https://www.eia.gov/electricity/monthly/

\bibitem{ref8} S. Pahwa, C. Scoglio, A. Scala, \textit{Abruptness of cascade failures in power grids}, Sci. Rep. {\bf 4}, 3694 (2014).

\bibitem{yagan1}
O. Yagan, {\it Robustness of power systems under a democratic-fiber-bundle-like model}, Phys. Rev. E {\bf 91}, 062811 (2015).

\bibitem{yagan2}
Y. Zhang, O. Yagan, {\it Optimizing the robustness of electrical power systems against cascading failures}, Sci. Rep. {\bf 6}, 27625 (2016). 

\bibitem{eq_long}
I. G. Main, {\it Damage mechanics with long-range interactions:
correlation between the seismic {\it b}-value and the fractal two-point correlation dimension},
Geophys. J. Int. {\bf 111}, 531 (1992).

\bibitem{power_long}
L. Shalalfeh, P. Bogdan, E. Jonckheere, {\it Evidence of long-range dependence in power grid}, in {\it 2016 IEEE Power and Energy Society 
General Meeting}, pages 1-5, 2016.

\bibitem{sadd}
M. Sadd, {\it Elasticity: Theory, applications and numerics, 3rd edition}, Elsevier, 2014.

\bibitem{alava}
M. J. Alava, P. K. V. V. Nukala, S. Zapperi, {\it Statistical models of fracture},
Adv. Phys. {\bf 55}, 349 (2006).

\bibitem{remote}
D. P. Hill, S. G. Prejean, {\it Dynamic triggering}, Treatise on Geophysics {\bf 4}, 257 (2007).

\bibitem{96outage}
D. N. Kosterev,  C. W. Taylor, W. Mittelstadt, {\it Model validation for the August 10, 1996 WSCC system outage}, IEEE Trans. Power Syst. {\bf 14},
967 (1999).

\bibitem{kawamura1}
H. Kawamura, T. Yamamoto, T. Kotani, H. Yoshino, {\it Asperity characteristics of the Olami-Feder-Christensen model of
earthquakes}, Phys. Rev. E {\bf 81}, 031119 (2010).

\bibitem{kawamura2}
Y. Ueda, S. Morimoto, S. Kakui, T. Yamamoto, H. Kawamura, {\it Dynamics of earthquake nucleation
process represented by the Burridge-Knopoff model}, Eur. Phys. J. B {\bf 88}, 235 (2015).

\bibitem{newmann05}
M. E. J. Newmann, {\it Power laws, Pareto distribution and Zipf's law}, Comp. Phys. {\bf 46}, 323 (2005).

\bibitem{pradhan}
S. Pradhan, A. Hansen, {\it Failure properties of loaded fiber bundles having a lower cut-off in fiber threshold distribution},
Phys. Rev. E {\bf 72}, 026111 (2005).

\bibitem{MLEClauset}
A. Clauset, C. R. Shalizi, M. E. J. Newman, {\it Power-law distributions of empirical data}, SIAM Rev. {\bf 51}, 661 (2009).

\bibitem{viola}
A. Levina, V. Priesemann, {\it Subsampling scaling}, Nat. Comm. {\bf 8}, 15140 (2017).

\bibitem{mao}
A. Mao, C. G. A. Harrison, T. H. Dixon, {\it Noise in GPS coordinate time series}, J. Geophys. Res. {\bf 104}, 2797 (1999).

\bibitem{behr}
J. Zhang, Y. Bock, H. Johnson, P. Fang, S. Williams, J. Genrich, S. Wdowinski, J. Berh, {\it South California GPS Geodetic array: Error analysis of daily position estimates and site velocities},
J. Geophys. Res. {\bf 102}, 5005 (1997).

\end{thebibliography}
\end{document}